\documentclass[aps,prl,twocolumn,showpacs,preprintnumbers,amsmath,amssymb]{revtex4-1}

 \usepackage{graphicx}

\begin{document}

\title{Two-dimensional Rydberg gases and the quantum hard squares model}
\author{S. Ji}
\author{C. Ates}
\author{I. Lesanovsky}
\affiliation{Midlands Ultracold Atom Research Centre (MUARC), School of Physics and Astronomy, The University of Nottingham, Nottingham, NG7 2RD, United Kingdom}

\begin{abstract}
We study a two-dimensional lattice gas of atoms that are photo-excited to high-lying Rydberg states in which they interact via the van-der-Waals interaction. We explore the regime of dominant nearest neighbor interaction where this system is intimately connected to a quantum version of Baxter's hard squares model. We show that the strongly correlated ground state of the Rydberg gas can be analytically described by a projected entangled pair state that constitutes the ground state of the quantum hard squares model. This correspondence allows us to identify a first order phase boundary where the Rydberg gas undergoes a transition from a disordered (liquid) phase to an ordered (solid) phase.
\end{abstract}

\pacs{67.85.-d, 32.80.Ee, 75.10.Kt, 05.30.Rt}

\maketitle
\begin{figure}
\centering
\includegraphics[width=\columnwidth]{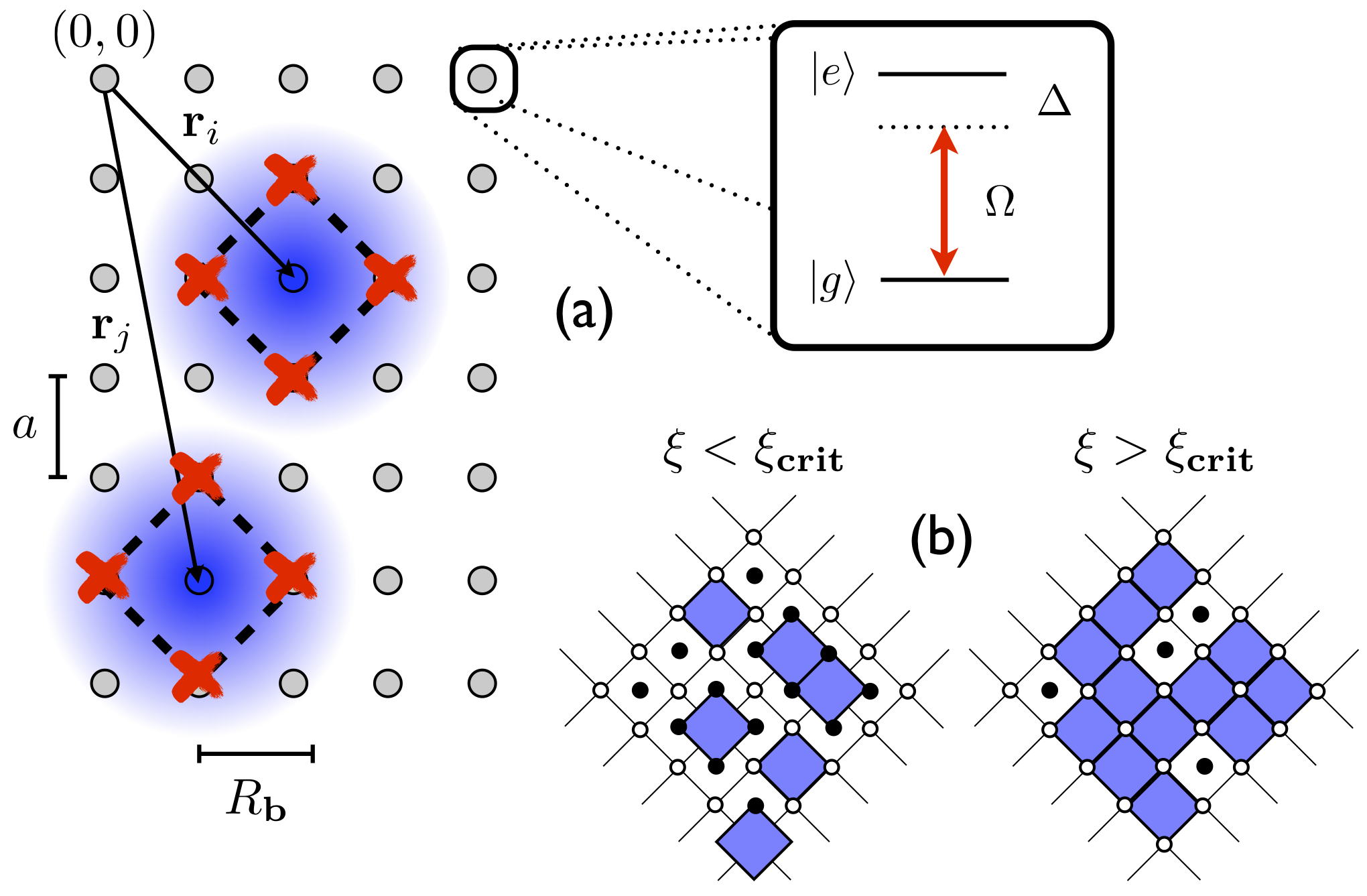}
\caption{(Color online) (a) Atoms arranged on a square lattice and coherently photo-excited from their ground state $| g \rangle$ to a Rydberg state $| e \rangle$ with Rabi frequency $\Omega$ and detuning $\Delta$. In the regime of ideal nearest-neighbor blockade a Rydberg atom entirely blocks the excitation of its four nearest neighbors (crosses). Exciting a Rydberg atom is, thus, equivalent to placing a hard square (dashed lines) centered on the position of the Rydberg atoms on the lattice. The shaded area indicates the range of the Rydberg-Rydberg interaction. (b) The phases of the quantum hard squares model [eq. (\ref{model_Hamil})] for different values of the parameter $\xi$ defined in the text. For $\xi < \xi_{\text{crit}}$: liquid phase in which both sub-lattices  (solid and open circles) are populated. For $\xi > \xi_{\text{crit}}$: ordered phase with broken sub-lattice symmetry.}
\label{system}
\end{figure}

The study of quantum many-body models with long-range interactions is a rapidly growing and flourishing field \cite{mibr+:06,blda+:08,sawa+:10} fueled by recent successes in creating ultra cold polar molecules \cite{nios+:08, degr+:08} as well as strongly interacting ultra cold gases of Rydberg atoms \cite{hera+:07,rabe+:08}. Rydberg gases offer widely tunable inter-atomic interactions which allow to explore correlated quantum matter in vastly different regimes, ranging from supersolids to exotic spin models \cite{welo+:08,howe+:10,pode+:10,hena+:10,pumi+:10,olgo+:10,leol+:10,webu:10,sepu+:11,olmu+:10,le:11}.
A particularly interesting regime is achieved at densities where the strong interaction between Rydberg states competes with the excitation laser. In the extreme case which is illustrated in Fig. \ref{system}(a), the simultaneous excitation of atoms located within a spatial radius $R_{\text{b}}$ (called the ``blockade radius'') is forbidden \cite{lufl+:01}. This blockade effect renders Rydberg atoms into ``hard objects'' whose mutual exclusion induces strong correlations in the ground state. In a one-dimensional setup this was analyzed in Ref. \cite{le:11}, where it was shown that the ground state can even be obtained analytically for certain combinations of the laser parameters.
A current challenge is to understand and to identify interesting and useful features of strongly correlated two-dimensional quantum systems.
Since the situation in two-dimensions is substantially more involved, exact analytical solution of strongly correlated two-dimensional quantum systems are scarce.

In this work we present a detailed study of a two-dimensional spin lattice gas motivated by recent experimental achievements in creating Rydberg lattice systems \cite{viba+:11_p} and the future perspective to implement quantum spin systems with polar molecules \cite{mibr+:06}. Our system is intimately related to a quantum version of a hard objects model - Baxter's hard squares \cite{ba:82} - whose ground state represents a coherent spin state with built-in nearest neighbor exclusion and belongs to the class of projected entangled pair states
(PEPS). We show that for certain parameters this strongly correlated state provides an accurate analytical description of the two-dimensional Rydberg gas. This highlights the possibility of creating PEPS in a physical system with binary interactions which is important as such states are considered a valuable resource for quantum information processing \cite{scwo+:07,chze+:09}. Furthermore, our study yields insights into the phase structure of the Rydberg gas allowing us to identify a first order quantum phase transition from a disordered quantum liquid to an ordered state [Fig. \ref{system}(b)].

The system under consideration consists of an ensemble of atoms regularly arranged on a square lattice and coherently photo-excited from their ground to a Rydberg $nS$-state by a laser with Rabi frequency $\Omega$ and detuning $\Delta$ [Fig. \ref{system}(a)]. We focus on the situation where the system is prepared in a Mott insulating state with one atom per site in a deep large-spacing optical lattice \cite{neli+:07}. The interaction between ground state atoms at different sites is negligible, while Rydberg atoms strongly interact with each other via a van-der-Waals interaction $U_{ij} = C_6 / | \mathbf{r}_i -\mathbf{r}_j|^6$. Here, $\mathbf{r}_i$, $\mathbf{r}_j$ denote the position vectors of the excited atoms, $C_6$ is the interaction coefficient and we define  the interaction strength at nearest neighbor distance $a$ as $V \equiv C_6/a^6$. In the following, we will consider the regime $R_{\text{b}} \gtrsim a$ shown in Fig. \ref{system}a, where the interaction is so strong that the simultaneous excitation of nearest neighbor particles is completely blocked. As demonstrated in \cite{le:11}, it is then convenient to work in the interaction picture with respect to the nearest neighbor interaction, since the Hamiltonian in this picture explicitly displays the effect of the excitation blockade.

Neglecting small corrections of order $\Omega^2/V$ the Hamiltonian  reads,
\begin{eqnarray}
H &=&\Omega \sum_{k,m}  \sigma^x_{k,m} \mathcal{P}_{k,m } + \Delta \sum_{k,m} n_{k,m} + \nonumber \\
 & & {}+ \frac{V}{8} \sum_{k,m} \left( n_{k,m} n_{k+1,m-1}  + n_{k,m} n_{k+1,m+1} \right) \, .
 \label{Rydberg_Hamil}
\end{eqnarray}
The Pauli matrix $\sigma^x_{k,m}$ changes the state of the lattice site $(k,m) \equiv \mathbf{r}_i/a$ from down to up and vice versa with the identification $\left|\uparrow\right>\equiv\left|e\right>$ and $\left|\downarrow\right>\equiv\left|g\right>$. The projector $n_{k,m} = 0,1$ determines if the site is occupied by a Rydberg atom, and we use its complement, $P_{k,m} = 1 - n_{k,m}$, to define a plaquette operator $\mathcal{P}_{k,m} \equiv P_{k-1,m} P_{k,m-1} P_{k+1,m} P_{k,m+1}$.

In the following, we will refer to the terms of $H$ that solely depend on the density $n_{k,m}$ as ``classical terms''. They describe the energy penalty to be paid if the laser is detuned by $\Delta$ from resonance, and the interaction between Rydberg atoms, where we have cut off the long-range tail of $U_{ij}$ at next-nearest neighbors, since its strength quickly decreases with increasing distance of the particles. In contrast, we will call the first term of eq.(\ref{Rydberg_Hamil}) ``quantum term''. It describes the coherent (de-)excitation of atoms at a rate $\Omega$ and is a five-particle interaction term: A particle at site $(k,m)$ can only be excited, if all of its four nearest neighbors are simultaneously in the ground state.  As indicated in Fig. \ref{system}(a), the action of the quantum term corresponds to placing a hard square on the site of the excitation or removing it from that site. Therefore, the realization of a two-dimensional Rydberg lattice gas in the ideal nearest neighbor blockade regime is equivalent to implementing a quantum lattice model of interacting hard squares which are created and annihilated by the laser.

The many-body ground state of Hamiltonian (\ref{Rydberg_Hamil}) is not known analytically. Its numerical determination by a straightforward diagonalization of $H$ is restricted to small system sizes, due to the exponential growth of the Hilbert space dimension with increasing particle number. We therefore propose an alternative method to study the ground state of the strongly interacting Rydberg lattice gas. The main idea is to consider an auxiliary model system whose many-body ground state wave function is known analytically and which possesses the same symmetry properties and the same quantum term as $H$. Using this wave function as a variational input state it is then possible to approximately determine the phase diagram of our original problem.

Our candidate model system is a quantum version of Baxter's classical hard squares model \cite{ba:82} on a square lattice. Its Hamiltonian is of the so called Rokhsar-Kivelson type \cite{cach+:05}, i.e. it can be written in the form,
\begin{equation}
H_{\text{HS}} =
\Omega \sum_{k,m} h^{\dagger}_{k,m} h_{k,m},
\label{model_Hamil}
\end{equation}
with positive-semidefinite, self-adjoint operators,
\begin{equation}
h_{k,m} = \sqrt{ \frac{1}{\xi^{-1} + \xi} }\left[
\sigma_{k,m}^x + \xi^{-1} n_{k,m} + \xi (1 - n_{k,m})
\right]  \mathcal{P}_{k,m} \, ,
\label{h_operator}
\end{equation}
and a non-negative parameter $\xi$.
The mathematical properties of the $h_{k,m}$ entail that the ground state energy of  $H_{\text{HS}}$ is zero and, consequently, the ground state wavefunction  $|\xi \rangle$ is annihilated by all $h_{k,m}$. It is given by
\begin{equation}
| \xi \rangle = \frac{1}{\sqrt{Z_{\xi}}} \exp \left( - \xi
\sum_{k,m} \,  \sigma_{k,m}^+ \mathcal{P}_{k,m}
\right) | 0 \rangle \, ,
\label{exact_wavefct}
\end{equation}
where $| 0 \rangle$ denotes the empty lattice (all atoms in the ground state), $\sigma_{k,m}^+ = (\sigma_{k,m}^x + i \sigma_{k,m}^y)/2$ is the spin raising operator at site $(k,m)$ and $Z_{\xi}$ the normalization constant. Physically, the many-body ground state of $H_{\text{HS}}$ is a coherent superposition of all many-particle states in which  Rydberg atoms are arranged in configurations compatible with the nearest neighbor (or hard squares) exclusion. Each configuration is weighted by a factor $(-\xi)^n$, where $n$ is the number of occupied lattice sites. The state $| \xi \rangle$ is a PEPS \cite{vewo+:06} and can be seen as a coherent spin state with built-in nearest-neighbor exclusion.

The normalization constant $Z_{\xi}$ is the partition function of the classical hard-squares model with fugacity $\xi^2$. Hence, the parameter $\xi$ controls the occupation of the lattice.  Using the partition function, classical observables like the density of particles and density-density correlations can be calculated. Interestingly, the expectation value of non-classical observables can also be calculated using $Z_{\xi}$, since they are connected to expectation values of classical observables taken in the ground-state $|\xi \rangle$. For example, the mean values of $\sigma_{k,m}^x$ is given by $\langle \xi | \sigma_{k,m}^x| \xi \rangle \equiv \langle \sigma_{k,m}^x \rangle_{\xi} = -2\; \xi^{-1} \langle n_{k,m} \rangle_{\xi}$.

The classical hard squares model displays an order-disorder phase transition between a  liquid and a solid phase with broken sub-lattice symmetry \cite{ba:82}. The thermal fluctuations that drive this transition are mapped  to quantum fluctuations by the PEPS solution (\ref{exact_wavefct}). Therefore, the quantum hard squares model will also undergo a liquid-solid  transition  at a critical value $\xi_{\text{crit}}\approx 1.9$. Here however, it is driven by quantum fluctuations \cite{vewo+:06} [c.f. Fig. \ref{system}(b)]. Since a order-disorder quantum phase transition is also anticipated for ultra cold Rydberg gases \cite{welo+:08,hena+:10,pumi+:10,webu:10,sepu+:11} and since the Hamiltonians (\ref{Rydberg_Hamil}) and (\ref{model_Hamil}) have the same symmetry properties, the state (\ref{exact_wavefct}) appears to be a natural candidate for performing a variational study of the phase diagram of the Rydberg lattice gas.

To solidify this statement we will now quantify the intimate connection of the Rydberg lattice gas problem with the quantum hard squares model. By multiplying out the quantum hard squares Hamiltonian (\ref{model_Hamil}) and comparing it termwise to eq. (\ref{Rydberg_Hamil}), one finds that the quantum term is identical in both cases. The classical terms of $H_\mathrm{HS}$ that are linear in $n_{k,m}$ are identical to those of $H$ if we choose the laser detuning such that $\Delta(\xi) = \Omega (\xi^{-1} - 5\xi)$. In addition we can make some of the two-body terms equal by choosing $V(\xi) = 8\Omega\, \xi$. These two choices define a manifold ($\xi$-manifold) in parameter space $(\Delta(\xi),V(\xi))$ along which the Hamiltonian (\ref{Rydberg_Hamil}) becomes $H(\xi) = H[\Omega,\Delta(\xi),V(\xi)]$. Note that along the $\xi$-manifold the Rydberg Hamiltonian and $H_\mathrm{HS}$ are similar but not equal, i.e. $H_{\text{HS}} = H(\xi) + \Delta H$.  The difference operator $\Delta H$ contains binary interactions between Rydberg atoms at distance $2a$ as well as three and four-body interaction terms that are all completely absent in the Rydberg Hamiltonian and, thus, cannot be accounted for by an appropriate parametrization.

\begin{figure}
\centering
\includegraphics[width=0.80\columnwidth]{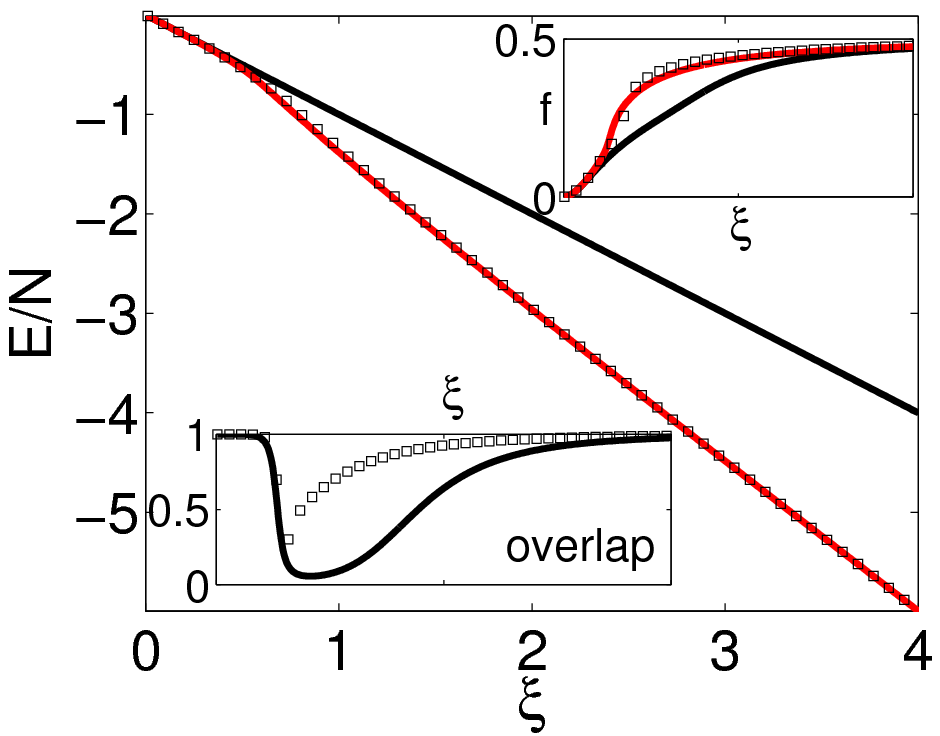}
\caption{(Color online) Main plot: Energy of the $6 \times 6$ Rydberg lattice gas on the $\xi$-manifold (c.f. red dashed line in Fig. \ref{phase_diagram}). Red line: exact ground state energy of (\ref{Rydberg_Hamil}). Black line: energy calculated using the coherent state (\ref{exact_wavefct}) with the same parameter $\xi$ as taken for the parameterization of the $\xi$-manifold. Squares:  Energy calculated by taking the PEPS with an independent parameter $\eta$ that minimizes $\langle \eta | H(\xi) | \eta \rangle$. Upper inset: Same for the fraction of excited atoms.  Lower inset: Overlap between the exact ground state of Hamiltonian (\ref{Rydberg_Hamil}) with the PEPS (\ref{exact_wavefct}) using the parameter $\xi$ (black) and the optimized parameter $\eta$ (squares). For details see text.}
\label{compare}
\end{figure}
To see how close nevertheless the coherent state (\ref{exact_wavefct}) resembles the true many-body ground-state $| G(\xi) \rangle$ of $H$ on the  $\xi$-manifold, we compare the numerically determined ground-state energy $E =\langle G(\xi) |H (\xi ) | G( \xi) \rangle$, as well as the average fraction of excited atoms $f =N^{-1} \sum_{k,m} \langle G(\xi) | n_{k,m}| G(\xi) \rangle$ with the energy and excitation fraction of the Rydberg gas calculated with $| \xi \rangle$. In addition, we compute the overlap $|\langle G(\xi) | \xi \rangle |^2$ between the exact ground state and hard squares coherent state. Fig. \ref{compare} shows the data for a $6\times 6$ lattice with periodic boundaries. For small $\xi$, i.e. low Rydberg density, both wave functions yield identical energies and lattice occupancies. This agreement is also reflected in the overlap integral, which shows that the many-body ground state of the Rydberg gas in the low density regime is given \emph{exactly} by a PEPS of the form (\ref{exact_wavefct}). In the opposite limit, $\xi \to \infty$, the  energies differ significantly, due to the difference $\Delta H$ of quantum hard squares and Rydberg gas Hamiltonian. However, we observe reasonable agreement of the calculated densities as well as in the overlap. This is because for large $\xi$ the ground state energy of both  the quantum hard-squares model as well as of the Rydberg gas is minimized by the configuration with maximal occupation of the lattice with hard squares. For intermediate $\xi$, however, both calculations give different results for the fraction of excited atoms and the overlap integral is small.

The remarkable point is that the exact ground state $| G(\xi) \rangle$ of $H(\xi)$ is still well approximated for any $\xi$ by a coherent state of the form (\ref{exact_wavefct}). To see this, we introduce an independent, variational parameter $\eta$, write the PEPS in terms of this parameter and optimize $|\eta \rangle$ by minimizing the energy functional $\langle \eta | H (\xi) | \eta \rangle$. To compute the functional, we express it in terms of expectation values of classical operators and numerically calculate them via the partition function $Z_{\eta}$ using a transfer matrix method \cite{gubl:02}. Comparing the exact ground-state energy with the variationally obtained one (Fig. \ref{compare}) we see excellent agreement for all $\xi$ as well as a very good agreement for the fraction of excited atoms. The simple optimization procedure also drastically improves the overlap, which is well above 0.5 for most values of $\xi$ and never drops below 0.25 which is remarkable for a many-body state in a Hilbert space of dimension $67022$ \cite{foot}. This result suggests that the 2D Rydberg gas permits the preparation of a PEPS of the form (\ref{exact_wavefct}) and that the hard squares coherent state (\ref{exact_wavefct}) is well suited for carrying out a variational study of the phase diagram of the Rydberg lattice gas.
\begin{figure}
\centering
\includegraphics[width=0.95\columnwidth]{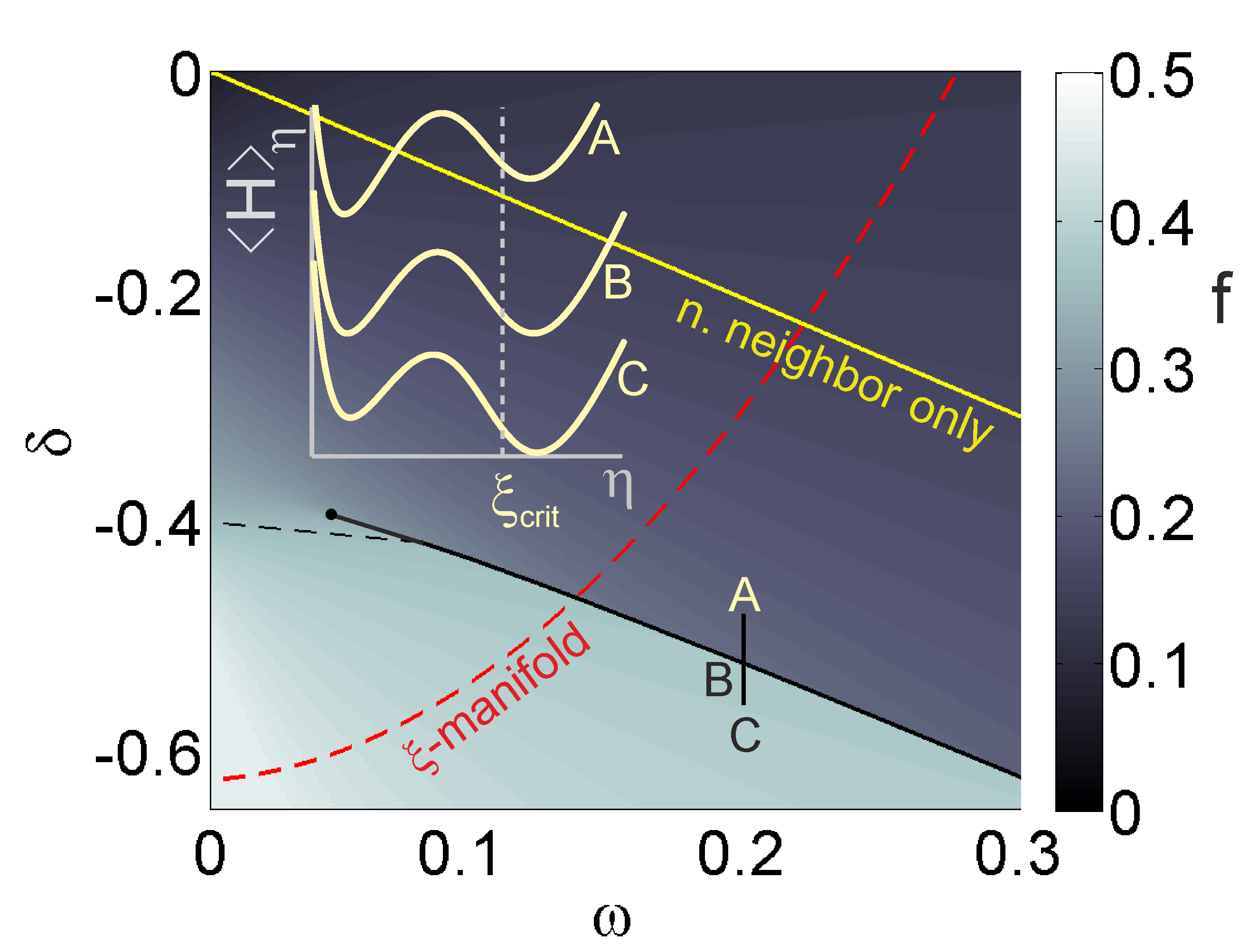}
\caption{(Color online) Variational phase diagram of the 2D Rydberg lattice gas with nearest-neighbor blockade on a $16 \times 16$ square lattice with periodic boundaries as function of the scaled detuning $\delta=\Delta/V$ and Rabi frequency $\omega=\Omega/V$. The dashed red line depicts the $\xi$-manifold along which the data of Fig. \ref{compare} have been determined and the dashed black line the values where $\eta_{\text{min}} = \xi_{\text{crit}}$ for $\omega < 0.1$. All other features are explained in the text.}
\label{phase_diagram}
\end{figure}

To determine the phase diagram of the Rydberg gas in the whole parameter range $(\Omega,\Delta,V)$ we minimize $\langle H \rangle_{\eta} \equiv \langle \eta | H | \eta \rangle$ with $H$ given by eq. (\ref{Rydberg_Hamil}). The position at which the functional becomes minimal is denoted by $\eta_\mathrm{min}$. The system is in the disordered (liquid) phase, if $\eta_\mathrm{min} < \xi_\mathrm{crit}$, where $\xi_\mathrm{crit}$ is the transition point of the hard squares model. For $\eta_\mathrm{min} > \xi_\mathrm{crit}$ it is in the (antiferromagnetically) ordered phase with broken sub-lattice symmetry [c.f. Fig.\ref{system}(b)].

The results of our variational approach are summarized in the phase diagram shown in Fig. \ref{phase_diagram} as a function of the scaled detuning $\delta = \Delta/V$ and laser driving $\omega = \Omega/V$.  In the region $\omega > \omega_1 \approx 0.1$, the system exhibits two distinct phases, a liquid and a solid phase. At large negative detuning, the Rydberg gas favors the solid phase near half filling ($f \lesssim 0.5$). The laser driving acts against the tendency of the atoms to order, such that if $|\delta|$ is reduced  the system undergoes a first order phase transition from the ordered phase to the liquid state (solid black line in Fig. \ref{phase_diagram}). Near the phase boundary the energy functional $\langle H \rangle_{\eta}$ has a double well structure with the minima located on either side of $\xi_\mathrm{crit}$. The transition is discontinuous, as the minima change their relative heights when passing through the transition (c.f. inset of Fig. \ref{phase_diagram}, which exemplifies this behavior along the line $A-B-C$).
For $\omega_1 > \omega > \omega_2 \approx 0.07$, we observe a qualitative change of the system's behavior.
Close to the phase boundary, the energy functional still has two distinct minima, however both are located below $\xi_{\text{crit}}$, indicating a first order transition from a low- to an high-density phase when reducing the detuning below $|\delta| \approx 0.4$. At $\omega = \omega_2$ the transition line appears to terminate at a critical point. Here the two wells of the variational energy $\langle H \rangle_{\eta}$ merge into a single minimum. Below $\omega =\omega_2$ our variational state $| \eta \rangle$ predicts a continuous crossover from the liquid to the half filled state with increasing negative detuning. In the limit $\omega \to 0$ we expect two classically ordered phases, one at quarter filling for $ 0 > \delta > - 0.5$ and the staggered phase at half filling for $-0.5 > \delta$. The variational state $|\eta \rangle$  cannot resolve these expected discontinuous transitions  for $\omega \ll \omega_2$. We emphasize that the discussed features of the phase diagram are a consequence of the long-range (beyond nearest neighbor) Rydberg interaction. For strict nearest neighbor interaction the variational calculation predicts a significantly shifted transition line and a first order phase transition from the empty lattice to a fully crystalline state starting at $\delta = \omega =0$ (c.f. yellow line in Fig. \ref{phase_diagram}). We have checked that including interaction terms that go beyond next-nearest neighbors only marginally change the phase diagram.

Our method for exploring the strongly correlated ground state of a 2D Rydberg gas appears to be fruitful also for other lattice geometries. Aside from studying the phase diagram it will be interesting to further explore the resulting PEPS, e.g., with focus on their power to perform computational tasks \cite{scwo+:07,vewo+:06}.

We acknowledge funding through EPSRC and discussions with J. P. Garrahan, B. Olmos and P. Kr\"uger.


\begin{thebibliography}{10}
\bibitem{foot} This reduction from an original dimension of $2^{36}$ is achieved by exploiting the symmetries of $H$.
\end{thebibliography}
%
\end{document}